\documentstyle[aps,prl,psfig,rotate]{revtex}

\tighten
\begin{document}
\preprint{\today}
\title{The incommensurate charge-density-wave instability
in the extended three-band Hubbard model}
\author{F. Becca, F. Bucci, and M. Grilli}
\address{Istituto Nazionale di Fisica della Materia and
Dipartimento di Fisica, Universit\`a di Roma ``La Sapienza'',\\
Piazzale A. Moro 2, Roma, Italy 00185}
\maketitle

\begin{abstract}
The infinite-$U$ three-band Hubbard model is considered in order to describe
the ${\rm CuO_2}$ planes of the high temperature superconducting
cuprates. The charge instabilities are investigated
when the model is extended with 
a nearest-neighbor repulsion between holes on copper $d$ and
oxygen $p$ orbitals and in the presence of a long-range
Coulombic repulsion. It is found that a first-order 
valence instability line ending with a
critical point is present like in the 
previously investigated model without long-range forces. However,
the dominant critical instability is the formation of
 incommensurate charge-density-waves, which always occur before the
valence-instability critical point is reached. 
An effective singular attraction arises in the proximity of 
the charge-density wave instability, accounting for 
both a strong pairing mechanism  and  for the
anomalous normal state properties.
\end{abstract}



\section{Introduction}

The crucial structural elements, present in all the superconducting cuprates,
are the ${\rm CuO_{2}}$ planes.
At half-filling, i.e. one hole per unit cell in the ${\rm CuO_{2}}$ 
planes, these 
materials are antiferromagnetic charge-transfer insulators 
and the holes mostly occupy the copper sites.  
The holes added by doping reside on the ${\rm CuO_{2}}$ planes and have 
a large amplitude on oxigen, thus showing that 
the strong hole-hole repulsion on copper is a relevant feature.
 Upon doping the system is driven toward 
a paramagnetic metallic phase which becomes 
superconducting at low temperature.
The metallic phase above $T_c$ presents many anomalous features,
which contrast with the usual behavior of the normal Fermi liquids.

The failure of Landau Fermi-liquid theory in the metallic phase
of the cuprates has been
ascribed to singular interactions arising in the proximity of some
critical point at zero temperature (quantum critical point, QCP)
\cite{MMP,pines,varma,tobepub,sing,perali}. More specifically
the complex behavior of these systems was recently interpreted 
in terms of a proximity to an incommensurate charge-density-wave (ICDW)
transition \cite{sing,perali,cina,evora} located at zero temperature near
the optimal doping. In the quantum critical region above it
no energy scale besides temperature rules the physics and 
strong critical fluctuations are  responsible for both non-Fermi liquid
behavior and large superconducting critical temperatures.
As soon as superconductivity takes place the CDW instability is hindered
and can only be recovered by destroying
the superconducting coherent state, like, e.g., 
in transport experiments under strong magnetic fields 
\cite{boebinger}. In the underdoped compounds, 
instead the instability would occur at finite temperature
were it not for the quenching due to superconducting local fluctuations
which can give rise to the 
appearance of charge and spin gaps of $d$-wave
symmetry as experimentally found 
\cite{marshall,harris,campuzano,rossat,jullien}
at a temperature $T^*$ above $T_c$. According to this proposal
\cite{cina,evora}, the underlying 
hindered charge instability provides the temperature dependent
pairing potential needed to explain the high crossover temperature $T^*$
of the gap formation and the peculiar doping dependence of $T^*$
(strongly increasing with decreasing doping, 
\cite{marshall,harris,campuzano}) with respect to the
value of the charge gap at T=0 (nearly constant with decreasing
doping, \cite{harris,loram}). 

The existence of a QCP  near optimal doping
is supported by several experimental findings. In particular
recent transport measurements in ${\rm La_{2-x}Sr_{x}CuO_{4}}$ (LSCO) under
strong magnetic field \cite{boebinger} investigated the normal phase of these
systems when superconductivity is suppressed. This analysis 
shows the existance of a QCP near optimal
doping. Indications in this sense
 are also provided by neutron scattering\cite{aeppli}
and by  the qualitative changes of
behavior at optimal doping detected by 
optical spectroscopy\cite{puchkov}, NMR \cite{jullien}, 
susceptibility \cite{batlogg},
neutron scattering \cite{rossat}, 
photoemission \cite{marshall,harris,campuzano},
specific heat \cite{loram}, thermoelectric power \cite{zhou},
Hall coefficient \cite{hwang}, 
resistivity \cite{boebinger,batlogg,ito}.
It is also suggestive that several quantities
(resistivity, Hall number, uniform susceptibility) display
a scaling behavior with a typical energy scale, which 
vanishes at optimal doping \cite{johnston,nakano,wuyts}.

Many indications exist that the above QCP involves charge ordering.
 Generically, the sizable doping at which the QCP occurs
suggests that  charge degrees of freedom are substantially
involved in the ordering phenomenon.
A direct observation of charge-driven ordering
was possible by neutron scattering
\cite{tranquada1,tranquada2,tranquada3},
in ${\rm La_{1.48}Nd_{0.4}Sr_{0.12}CuO_4}$ 
where the related Bragg peaks were detected. For this specific
compound the low temperature tetragonal lattice structure 
pins the CDW and gives static order and 
semiconducting behavior (see also the case of 
${\rm La_{1.88}Ba_{0.12}CuO_4}$).
Increasing the Sr content at fixed Nd concentration, the pinning
effect is reduced leading to metallic and superconducting behavior.
In this latter case, the existence of dynamical
ICDW fluctuations is suggested by the presence of dynamical incommensurate
spin scattering, although 
the charge peaks are too weak to be observed. 
In this regard, also the 
${\rm La_{2-x}Sr_xCuO_4}$ is expected to display dynamical charge
fluctuations with doping-dependent spatial modulation as
indeed observed in the magnetic scattering \cite{yamada}.
ICDW have been proposed from extended X-ray absorption
fine structure (EXAFS) experiments both in optimally
doped LSCO \cite{bianconi1} and ${\rm Bi_2Sr_2CaCu_2O_{8+x}}$
(Bi-2212) \cite{bianconi2}.
Superstructures have also been detected in Bi-2212 from 
X-ray diffraction \cite{bianconi3}. 

The occurrence of an ICDW-QCP
 is theoretically substantiated by explicit
findings within the single-band infinite-U Hubbard model in the presence of  
an Holstein 
electron-phonon coupling. In the absence of long-range (LR) coulombic forces
and for sizable but realistic electron-phonon coupling,
this model displays phase separation (PS) into macroscopically
large half-filled insulating regions and metallic hole-rich doped regions.
When the LR Coulomb repulsion between holes 
is included the  $q=0$ charge instability
is prevented. Nevertheless a phase-separation tendency remains in the
system and shows up on a local basis:
The PS region of the phase diagram is replaced 
at small doping by a smaller ICDW
instability region characterized by a non-vanishing  
incommensurate wavevector $q_{c}$.  

As discussed above, the occurrence of an incommensurate charge instability and 
of the related singular attractive interactions is a key
point, which could provide a unified explanation for both the normal and
superconducting properties of the cuprates. In this regard it seems
quite important to investigate how generic and robust is the above
scenario based on an ICDW instability arising from
the competition between PS and LR repulsion.

As far as PS is concerned, this is a common feature of strongly correlated
electron systems: It has been found in models with nearest-neighbour Coulomb 
interactions \cite{weak,coul1,coul2,coul3,ruckenstein}, 
in magnetic models \cite{tj1,tj2}, 
 and in models with phononic interactions \cite{longr,cg}. 
Indeed a strong on-site correlation 
drastically renormalizes the kinetic energy, which would tend to delocalize
the carriers into Bloch quasiparticle states. 
Then short-range interactions 
(magnetic, phononic, nearest-neighbor coulombic...)
introducing effective attractions between the carriers
may dominate and give rise to charge aggregation in highly doped 
metallic regions together with charge depletion in spatially separated 
(ordered) regions with no itinerant charges. 

As pointed out by Emery and Kivelson \cite{physc}
LR Coulomb forces effectively oppose the separation
of charged particles suppressing long-wavelenght density fluctuations.
This may lead to either dynamical slow density 
fluctuations \cite{physc} or static 
ICDW \cite{sing}, \cite{perali}, \cite{coul2}, \cite{emery}.
However,  whereas
the occurrence of PS is well established in many models of strongly
interacting electrons, the occurrence of ICDW from 
LR coulombic repulsion frustrating PS in a microscopic electron model
was only discussed in a single-band Hubbard-Holstein model 
\cite{sing,longr} and only
marginally considered in a three-band extended Hubbard model \cite{coul2}.
Then, although the physical ingredients seem fully general,
to gain further understanding of the ICDW-QCP scenario, it is 
desirable to investigate other theoretical models of strongly
interacting electrons in the presence of LR forces. In this way
one can both (a) explicitly check the generic character of the
physical mechanism and (b) highlight the specific features arising from
the various interactions. In this regard, the simple single-band
Hubbard-Holstein model having already been considered, the 
natural extension seems to be a multiband Hubbard model with
purely electronic interactions. 


The modelization of ${\rm CuO_{2}}$ planes
by  the three-band Hubbard model, in which the
copper and oxigen degrees of freedom are explicitly taken into account,
was proposed by Varma {\it et al.} together with the inclusion of the 
nearest-neighbor
Coulomb repulsion in order to stress the role 
of the charge-transfer (copper-oxigen) fluctuations \cite{eccitoni}.
Littlewood {\it et al.} \cite{little1},\cite{little2}, using a weak-coupling 
approach (for a finite not too large repulsion on copper site) have found 
that the presence of a nearest-neighbour Coulomb 
repulsion $V$ of order of the bandwidth can drive the system toward a 
valence instability (VI), characterized by strong charge fluctuations between
the copper $d$ orbitals and the oxigen $p$ orbitals. 
If the total number of particles is allowed 
to fluctuate, the VI has been shown \cite{weak} to be embedded in 
an unstable region where the system phase-separates. 
Also in a strong-coupling approach 
\cite{coul1}, \cite{coul2} the mean-field solution at fixed number of 
particle displays a VI line ending at the so-called  VI  point, where 
the charge-transfer excitonic mode completely softens. 
However, when the number of particles can fluctuate, this line 
is surrounded by a region of negative compressibility, i.e. by a 
region in which the system phase-separates.
Therefore PS always occurs both in weak and strong 
coupling before any VI takes place.

In this paper we address the problem of introducing the LR forces in a
three-band Hubbard model. In this way we study how the charge-transfer
fluctuations, which are able to provide a strong attraction in the
particle-hole channel and lead to PS in the pure
short-range case, are affected by a Coulomb potential.
 In section II we introduce the model. In particular
we work within a strong coupling approach: An 
infinite repulsion between holes on the same copper site
 is handled by means of a standard slave-boson technique. Fluctuations
around mean-field values are taken into account up to the gaussian order
using a systematic $1/N$ expansion. This section is rather technical
and the uninterested reader can skip it 
only retaining the form of the model provided by the pure short range terms
[Eq. (\ref{lamadreditutto}) 
with $U_d=\infty$] and by the LR potential  [Eqs. (\ref{cco})-(\ref{cupo})].

In section III we will show that, as in the LR Hubbard-Holstein model, the 
phase diagram of three-band Hubbard model in the presence of LR interactions
displays a finite-$q$ instability related to CDW formation.
We find a singular scattering between quasiparticles near the line on which 
the generalized density-density correlation functions diverge
at a finite $q_c$. By
contrast in the proximity of the  VI  line no singular scattering is found.
Only at the isolated VI  point the interaction takes a singular form due to
vanishing of the excitonic energy, but 
this has no physically relevant consequences
since the VI point is 
always inside the unstable region with respect to the ICDW formation.
This conclusion differs from the results obtained 
by Varma in a three-band Hubbard model \cite{varma,tobepub}, 
where an instability related to 
intracell charge-transfer currents was found, together with
singular interactions. However, in the model considered by Varma a relevant
role was played by general symmetry properties,
which are absent in our case, thus rendering a direct
comparison between the two models and results questionable.
 In section IV we discuss the
effective interaction between quasiparticles, whereas in section V we 
draw our conclusions.

\section{The model}

\subsection{Introduction of LR potential}

The Hamiltonian of the three-band Hubbard model in the presence of a
local repulsion on the copper site $U_d$
and a nearest-neighbour repulsion $V$ reads
\begin{eqnarray} 
H&=&\epsilon_{d}^{0} \sum_{i \sigma} {d_{i \sigma}^{\dag}}  {d_{i \sigma}} +
\epsilon_{p}^{0} \sum_{i \sigma \alpha=x,y } 
{p_{i \alpha \sigma}^{\dag}}  {p_{i \alpha \sigma}}
-t_{pd} \sum_{i \sigma \eta=\pm x,y} sgn (\eta) 
\left ( {d_{i \sigma}^{\dag}}  {p_{i \eta \sigma}} +h.c. \right ) \nonumber\\
&-&t_{pp} \sum_{i \sigma} p_{i+x \sigma}^{\dag} 
\left [ p_{i-y \sigma}-p_{i+y \sigma} 
+\left (i \rightarrow i+1 \right ) +h.c. \right ] 
+U_{d} \sum_{i} n_{i \uparrow}^{d} n_{i \downarrow}^{d} 
+V \sum_{i \sigma \sigma^{\prime}\eta=\pm x,y} n_{i \sigma}^{d} 
n_{i \sigma^{\prime}}^{\eta} 
\label{lamadreditutto}
\end{eqnarray}
Where ${d_{i \sigma}^{\dag}} $ (${d_{i \sigma}}$) creates (destroys) a hole 
in the copper site $i$, 
${p_{i \alpha \sigma}^{\dag}} $
(${p_{i \alpha \sigma}}$) in the oxigen site $i+\alpha$, 
$n_{i \sigma}^{d}={d_{i \sigma}^{\dag}}  {d_{i \sigma}}$, 
$n_{i \sigma}^{\eta}={p_{i \eta \sigma}^{\dag}}  {p_{i \eta \sigma}}$ and 
$\epsilon_{d}^{0}$ and $\epsilon_{p}^{0}$ 
are the bare energy levels. We work in hole
representation and we denote by $\delta$ the number of holes per unit cell
added by doping: $ \sum_{\sigma,\alpha=x,y} 
\left[\langle n_{i\sigma}^{\alpha}\rangle +
\langle n^d_{i\sigma} \rangle \right]= (1+\delta)$.
 
In the case of $U_{d}=\infty$ (strong-coupling) 
no double occupancy arises on copper sites
$\sum_{\sigma} {d_{i \sigma}^{\dag}}  {d_{i \sigma}} \le 1$.
We handle this constraint within the standard slave-boson technique
\cite{barnes,col,rn,millis,kotliu}
by introducing a new boson degree of freedom labeling the empty site
${d_{i \sigma}^{\dag}}  
\rightarrow {d_{i \sigma}^{\dag}}  b_{i}, 
\;\;\;\;\; {d_{i \sigma}} \rightarrow b_{i}^{\dag} {d_{i \sigma}}$.
Since the site can only be either singly occupied by a fermion or singly
occupied by a boson, the constraint becomes 
$\sum_{\sigma} {d_{i \sigma}^{\dag}}  {d_{i \sigma}} +b_{i}^{\dag} b_{i}=1$.
In order to avoid perturbative assumptions on the coupling constants we use
the large-$N$ expansion, assuming the spin index to run from $1$ to $N$ and
letting $N \rightarrow \infty$. So the constraint is relaxed into
\begin{equation}
\label{relax}
\sum_{\sigma} {d_{i \sigma}^{\dag}}  
{d_{i \sigma}} +b_{i}^{\dag} b_{i}=\frac{N}{2},
\end{equation}
showing  that $b_{i} \sim \sqrt{N}$.
Consistently we rescale the coupling
$t_{pd} \rightarrow t_{pd}/\sqrt{N}$, $V\rightarrow V/N$.
Then the partition function of the system can be written as a functional
integral
\begin{eqnarray}
{\cal Z}=\int {\cal D} d_{\sigma}^{\dag} {\cal D} d_{\sigma} 
{\cal D} p_{\alpha \sigma}^{\dag} {\cal D} p_{\alpha \sigma} 
{\cal D} b^{\dag} {\cal D} b {\cal D} \lambda {\cal D} X {\cal D} Y
{\rm exp} \left ( -\int_{0}^{\beta}d\tau S \right ),
\end{eqnarray}
with
\begin{equation}
S=\sum_{i \sigma} {d_{i \sigma}^{\dag}}  \frac{\partial {d_{i \sigma}}}
{\partial \tau}+
\sum_{i  \sigma \alpha=x,y} {p_{i \alpha \sigma}^{\dag}}  
\frac{\partial {p_{i \alpha \sigma}}}{\partial \tau}+
\sum_{i \sigma} b_{i}^{\dag} \frac{\partial b_{i}}{\partial \tau}
+i \sum_{i}\lambda_{i} \left ( b_{i}^{\dag} b_{i}-q_{0}N \right )
+\frac{N}{2V}\sum_{i} \left ( X_{i}^{2}+Y_{i}^{2} \right ) + {\cal H},
\label{action}
\end{equation}
\begin{eqnarray}
{\cal H}&=&\sum_{i \sigma} 
\left (\epsilon_{d}^{0}+i 
\lambda_{i}+X_{i}+iY_{i} \right ) {d_{i \sigma}^{\dag}}  {d_{i \sigma}} 
+\sum_{i \sigma \alpha=x,y } 
\left (\epsilon_{p}^{0}-X_{i}+iY_{i} 
\right ){p_{i \alpha \sigma}^{\dag}}  {p_{i \alpha \sigma}} \nonumber\\
&-&\frac{t_{pd}}{\sqrt{N}} \sum_{i \sigma \eta=\pm x,y} 
sgn (\eta) \left ( b_{i} {d_{i \sigma}^{\dag}}  
{p_{i \eta \sigma}} +h.c. \right ) 
-t_{pp} \sum_{i \sigma} p_{i+x \sigma}^{\dag} 
\left [ p_{i-y \sigma}-p_{i+y \sigma} 
+\left (i \rightarrow i+1 \right ) +h.c. \right ]. 
\label{short}
\end{eqnarray}
$\lambda$ is a Lagrange multiplier introduced to enforce 
the constraint in Eq. (\ref{relax}).
Writing the nearest-neighbour Coulomb interaction as
\[
\frac{V}{N}\sum_{i} n_{i}^{d} n_{i}^{p}=
\frac{V}{2N}\sum_{i} \left [ 
\left ( n_{i}^{d}+\frac{1}{2} n_{i}^{p} \right )^{2}-
\left ( n_{i}^{d}-\frac{1}{2} n_{i}^{p} \right )^{2} \right ].
\]
where $n_{i}^{p} =
\sum_{i \sigma \eta= \pm x,y} n_{i \sigma}^{\eta},\;n_{i}^{d} =
\sum_{i \sigma} n_{i \sigma}^{d}$, 
we have decoupled this interaction by means of an Hubbard-Stratanovich 
transformation introducing the fields $X_i$ and $Y_i$.
Note that in the leading order of the large-$N$ expansion only the Hartree
factorization is present, the Fock term being suppressed by a factor
$\frac{1}{N}$ due to the sum over the spin. In order to consider the Fock 
factorization also, one needs to add the term
\begin{equation}
\label{fock}
H_{V_{2}}=\frac{V_{2}}{N} \sum_{i \sigma \sigma^{\prime} \eta=\pm x,y}
{d_{i \sigma}^{\dag}}  
d_{i \sigma^{\prime}} p_{i \eta \sigma^{\prime}}^{\dag} {p_{i \eta \sigma}}.
\end{equation} 
This term can be decoupled by a complex Hubbard-Stratanovich field 
$Z_{i \eta}$. However we shall see in Appendix B
that the Fock term does not modify the properties 
of the model from a qualitative point of view. So we discard it from now on.

Now we extend the formalism in order to consider the Coulomb
LR potential by taking into account the
 symmetries of the underling lattice, that is of
the square lattice of copper atoms with a two-oxygen basis
along the directions $(a/2,0)$ and $(0,a/2)$.
Then we add to the hamiltonian the term:
\begin{equation}
\label{cco}
H_{coul}=\frac{1}{2N} \sum_{q \sigma \sigma^{\prime}} 
n_{q \sigma}^{a} \Phi^{a b}(q) n_{q \sigma}^{b},
\end{equation}
where $a,b=d,x,y$ and the sum over $a$ and $b$ is understood. 

$\Phi^{a b}$ reads as
\begin{equation}
\Phi^{a b}(q)= V_{c}(q)
\left (
\begin{array}{ccc}
1                                                                      & 
\cos \left (\frac{aq_x}{2} \right )                                    & 
\cos \left (\frac{aq_y}{2} \right )                                    \\
\cos \left (\frac{aq_x}{2} \right )                                    & 
1                                                                      & 
\cos \left (\frac{aq_x}{2} \right )\cos \left (\frac{aq_y}{2} \right ) \\
\cos \left (\frac{aq_y}{2} \right )                                    & 
\cos \left (\frac{aq_x}{2} \right )\cos \left (\frac{aq_y}{2} \right ) &
1
\end{array}
\right ),
\label{ccomatrix}
\end{equation}
and $V_{c}$ is given by
\begin{equation}
\label{cupo}
V_{c}(q)=
\frac{{\tilde V}}{\sqrt{ \left \{ 
\frac{\epsilon_{\parallel} b^{2}}{\epsilon_{\perp} a^{2}} \left [
\cos(aq_{x})+\cos(aq_{y})-2 \right ] -1 \right \}^{2}-1}},
\end{equation}
We define $\epsilon_{\parallel}$ and $\epsilon_{\perp}$ as
the dielectric constant in and out the Cu-O plane, $a$ and $b$ 
the in-plane and out-plane lattice constants; ${\tilde V}$ measures
the strength of the Coulomb potential. 

$V_{c}(q)$ is obtained through the solution of the $3D$ Laplace 
equation for a point-like charge in a cubic lattice, projected onto the 
$z=0$ plane \cite{longr}. The matrix form of $\Phi^{a b}(q)$  reproduces
the structure of the Fourier-space Laplacian operator in the orbital space.
In the small $q$ limit $V_{c}(q)$ has the right behavior
\[
\lim_{q \rightarrow 0} V_{c}(q) \sim \frac{1}{q}.
\]
corresponding to 
a two dimensional plane embedded in a three dimensional space.

The partition
function including the long-range term now reads
\begin{equation}
{\cal Z}=\int {\cal D} d_{\sigma}^{\dag} {\cal D} d_{\sigma} 
{\cal D} p_{\alpha \sigma}^{\dag} {\cal D} p_{\alpha \sigma} 
{\cal D} b^{\dag} {\cal D} b {\cal D} \lambda {\cal D} X {\cal D} Y
{\cal D} \Omega^{a} {\rm exp} \left ( -\int_{0}^{\beta}d\tau {\tilde S} 
\right ),
\label{long}
\end{equation}
\begin{equation}
{\tilde S} = S + \frac{N}{2}\sum_{q} 
\Omega_{q}^{a} \Phi_{a b}^{-1}(q) \Omega_{q}^{b} 
+i \sum_{q \sigma} \Omega_{q}^{a} n_{q \sigma}^{a},
\label{long2}
\end{equation}         
where $\Omega_{q}^{a}$ is a real vector field, introduced to 
decouple the LR Coulombic interaction. Notice that 
in Eq. (\ref{cupo}) we use the 
long-wavelength uniform values for the dielectric
constants in order to obtain the repulsive Coulomb potential
at long-distance. We have also included in the
Hamiltonian the explicit n-n Coulomb repulsion $V$. This term is needed to
describe the reduced screening present at short distances between 
localized charges, which leads to an enhanced repulsion between n-n sites.

\subsection{Large-$N$ Expansion} 

In this section we present the formalism needed to handle 
the gaussian fluctuations of the
boson fields around the saddle-point solutions. 
From now on we choose the radial gauge  \cite{rn}.
In this gauge
the phase of $b_{i}$ is gauged away and only the modulus
$r_{i}$ is considered, while $\lambda_{i}$ acquires a time dependence.
The Hamiltonian of the coupled fermions and bosons 
is then written in a
compact form, by introducing a seven-component field 
${\cal A}^{\mu}=(\delta r, \delta \lambda, \delta X, \delta Y, \Omega^{a})$ 
defined as the fluctuating part of the fields around 
the saddle-point solutions
\begin{eqnarray}
r_{i}&=&r_{0}(1+\delta r_{i})                    \nonumber\\
\lambda_{i}&=&-i \lambda_{0}+ \delta \lambda_{i} \nonumber\\
X_{i}&=&X_{0}+\delta X_{i}                       \nonumber\\
Y_{i}&=&-iY_{0}+\delta Y_{i}                       \nonumber\\
\Omega_{q}^{a}&=&\Omega_{q}^{a}                  \nonumber.  
\end{eqnarray}
We have not included the (infinite) 
zero-momentum component of the Coulombic field 
$\Omega_{q}^{a}$ since we are assuming that it 
is cancelled by the contribution of a uniform charged ionic background.

The Hamiltonian reads
\begin{equation}
{\cal H}={\cal H}_{MF}+{\cal H}_{bos}+{\cal H}_{int},
\end{equation}
where, using the notation $\Psi_{k \sigma \alpha}=(d_{k \sigma}, 
ip_{x k \sigma}, ip_{y k \sigma})$, ${\cal H}_{MF}$ is the mean-field 
fermion hamiltonian (in units of $a=1$)
\begin{equation}
\label{hhmf}               
{\cal H}_{MF}(k)= 
\left (
\begin{array}{ccc}
\epsilon_{d}                                       & 
-2r_{0}t_{pd} \sin \left (\frac{k_{x}}{2} \right ) & 
-2r_{0}t_{pd} \sin \left (\frac{k_{y}}{2} \right ) \\
-2r_{0}t_{pd} \sin \left (\frac{k_{x}}{2} \right ) & 
\epsilon_{p}                                       &
-2t_{pp} \beta_{k}                                 \\
-2r_{0}t_{pd} \sin \left (\frac{k_{y}}{2} \right ) & 
-2t_{pp} \beta_{k}                                 &
\epsilon_{p}                                                          
\end{array}
\right )
\end{equation}
where $\epsilon_{p}=\epsilon_{p}^0-X_0+Y_0$ and $\epsilon_{d}=
\epsilon_{d}^0+\lambda_0+X_0+Y_0$ are the renormalized energy oxigen and
copper levels respectively, and
$\beta_{k}=2 \sin \left (k_{x}/{2} \right) 
\sin \left (k_{y}/{2} \right)$.

The quasiparticle basis $\tilde{\Psi}_{k}$ is obtained from the
unitary transformation, $\tilde{\Psi}_{k}=U(k) \Psi_{k}$ which diagonalizes
the mean-field Hamiltonian $H_{MF}$, so that
\[            
{\cal H}_{MF}=
\sum_{k \sigma \alpha \beta} 
\Psi_{k \sigma \alpha}^{\dag} {\cal H}_{MF}^{\alpha \beta}(k) 
\Psi_{k \sigma \beta}=
\sum_{k \sigma \alpha} E_{\alpha}(k) \tilde{\Psi}_{k \sigma \alpha}^{\dag} 
\tilde{\Psi}_{k \sigma \alpha}.
\]
The boson-fermion interaction term can be written in the form
\begin{equation}
\label{int}
{\cal H}_{int}=
\sum_{k q \sigma} \Psi_{k+(q/2) \sigma}^{\dag} \Lambda^{\mu}(k,q)
\Psi_{k-(q/2) \sigma} {\cal A}^{\mu}(q)=  
\sum_{k q \sigma} \tilde{\Psi}_{k+(q/2) \sigma}^{\dag} 
\tilde{\Lambda}^{\mu}(k,q) \tilde{\Psi}_{k-(q/2) \sigma} {\cal A}^{\mu}(q).
\end{equation}
The $3 \times 3$ boson-fermion interaction 
vertices $\Lambda^{\mu}(k,q)$ in the
orbital operator basis can be obtained from Eq. (\ref{long2}) and are shown
in Appendix A.
The quasiparticle vertices $\tilde{\Lambda}^{\mu \nu}(k,q)$ are given
by
\begin{equation}
\label{verti}
\tilde{\Lambda}^{\mu}(k,q)=U \left (k+\frac{q}{2} \right )
\Lambda^{\mu}(k,q) U^{\dag} \left (k-\frac{q}{2} \right ).
\end{equation}
Finally the purely bosonic part of the hamiltonian is
\begin{equation}
{\cal H}_{bos}=N \sum_{q \mu \nu} {\cal A}^{\mu}(q) B^{\mu \nu}(q)
{\cal A}^{\nu}(-q),
\end{equation}
where the $7\times 7$ $B(q)$ matrix is given by:
\begin{equation}
B(q)= 
\left (
\begin{array}{cc}
B_{sr} & 0                        \\
0      & \frac{1}{2} \Phi^{-1}(q) 
\end{array}
\right ),
\end{equation}
and the only nonzero elements of $B_{sr}$ are
$B_{sr}^{11}=r_{0} \lambda_{0}$, $B_{sr}^{12}=B_{sr}^{21}=ir_{0}^{2}$ and 
$B_{sr}^{33}=B_{sr}^{44}=1/2V$. \\
At leading order in $1/N$,
the dressed propagator of the ${\cal A}^{\mu}$ field is
\begin{equation}
\label{prop}
D^{\mu \nu}(q, \omega)= \langle {\cal A}^{\mu}(q,\omega) 
{\cal A}^{\nu}(-q,-\omega) \rangle = 
\frac{1}{N} \left [ 2B(q)+\Pi(q,\omega) \right ]_{\mu \nu}^{-1},
\end{equation}
where $\Pi^{\mu \nu}(q,\omega)$ are the bare polarization bubbles:
\begin{equation}
\label{bubble}
\Pi^{\mu \nu}(q,\omega)=
\sum_{k \alpha \beta} 
\frac{f_{\alpha}(k+q/2)-f_{\beta}(k-q/2)}  
{E_{\alpha}(k+q/2)-E_{\beta}(k-q/2)-\omega}  
\tilde{\Lambda}_{\alpha \beta}^{\mu}(k,q)
\tilde{\Lambda}_{\beta \alpha}^{\nu}(k,-q).
\end{equation}
The factor $2$ multiplying the bare boson propagator is due to the 
fact that in the radial gauge the bosonic fields are real. 
Correspondingly, at leading order in $1/N$,
the density-density correlation functions are given by
\begin{equation}
\chi_{\alpha \beta}(q,\omega)=\frac{1}{N}
\sum_{\sigma \sigma^{\prime}} 
\langle n_{\alpha \sigma}(q)n_{\beta \sigma^{\prime}}(-q)\rangle=
\chi_{\alpha \beta}^{0}(q,\omega)+
\sum_{\mu \nu} P_{\alpha \mu}^{0}(q,\omega) D^{\mu \nu}
P_{\nu \beta}^{0}(q,\omega), 
\label{fdc}
\end{equation}
where
\begin{equation}
\label{barebl1}
\chi_{\alpha \beta}^{0}(q,\omega)=\frac{1}{N}
\sum_{\sigma \sigma^{\prime}} 
\langle n_{\alpha \sigma}(q)n_{\beta \sigma^{\prime}}(-q)\rangle_{0}
\end{equation}
are the bare density-density correlation functions, and
\begin{equation}
\label{barebl2}
P_{\alpha \mu}^{0}(q,\omega)=\frac{1}{N} 
\sum_{\sigma \sigma^{\prime}} \langle n_{\alpha \sigma}(q)
\sum_{k \gamma \delta} 
\Psi_{k \sigma^{\prime} \gamma}^{\dag} \Lambda_{\gamma \delta}^{\mu}(k,q)
\Psi_{k+q \sigma^{\prime} \delta} \rangle _{0},
\end{equation}
with $\alpha=d,p_{x},p_{y}$ and $\mu$ running on the boson indices. Linearly 
combining $\chi_{\alpha\beta}$, we calculate the
total density correlation function
\begin{equation}
\label{kappa}
\chi_{nn}(q,\omega)=\langle (n_{p}+n_{d})(n_{p}+n_{d}) \rangle
\end{equation}
and the charge-transfer correlation function 
\begin{equation}
\label{ct}
\chi_{CT}(q,\omega)=\langle (n_{p}-n_{d})(n_{p}-n_{d})\rangle.
\end{equation}

Within this formalism it is possible to calculate the 
residual (order $1/N$) effective scattering amplitude
between quasiparticles in the lowest band (where the Fermi level lays)
\begin{equation}
\label{effect}
\Gamma(k,k^{\prime}, \omega)=
-\tilde{\Lambda}_{11}^{\mu}(k^{\prime},-q)
D^{\mu \nu}(q,\omega)
\tilde{\Lambda}_{11}^{\nu}(k,q).
\end{equation}

We note in passing that, being the bare bubble defined in Eq. (\ref{bubble}) 
and the 
vertices defined in Eq. (\ref{verti}) non-singular functions of $q$, in 
the present RPA-like 
approximation the effective interaction Eq. (\ref{effect}) 
and the correlation functions Eq. (\ref{fdc}) 
always diverge together, since any
singularity may only arise from the boson propagator entering both quantities

\section{Phase Diagram and Correlation Functions}

The three-band extended Hubbard model with short-range interactions
has been widely studied in both
weak-coupling \cite{eccitoni,little1,little2,weak} and 
in semianalytic \cite{coul1,coul2,ruckenstein} or fully numeric 
\cite{gagliano,muramatsu,dopf1,dopf2,sudbo,sudbo2}
strong-coupling approaches. Before presenting the effects of LR forces
on this model, we recall some of the most important results for 
the pure short-range case.
In particular we summarize the phase-diagram in the $\Delta_0-\delta$ space, 
$\Delta_0=\epsilon^0_p-\epsilon^0_d$ being the bare difference
between the copper and oxygen atomic levels, for various given
values of the nearest-neighbor Cu-O repulsion $V$
 in the more realistic strong-coupling approach.
At half-filling a metal-charge-transfer-insulator (MCTI) transition is
present, for $\Delta_0$ larger than a critical value ($\Delta_0 +V> 
3.3t_{pd}$ when $t_{pp}=0$). Away from half-filling,
increasing $V$ the system shows phase-separation
above a critical value $V^{*}$ ($V^{*} \simeq 1.63t_{pd}$ for $t_{pp}=0$).
If $V$ is increased further, two solutions with 
different occupations of the copper and oxigen orbitals can be found
solving the self-consistent equations {\it at fixed number of 
particles}. Correspondingly to these solutions, 
the phase diagram, at finite doping, would 
display a ``first-order'' line between a d-like metal and a p-like metal 
ending in a ``second-order'' valence instability (VI) point
characterized by the softening of the energy of the $p-d$ 
exciton mode at $q=0$. However, 
if the number of particles is allowed to fluctuate 
the VI line is always inside a region of negative compressibility 
where the system phase-separates.  Then the  VI  point is 
not physically attainable. The region of negative
compressibility is delimited by a spinodal line on which the
compressibility diverges.
Together with the compressibility, all the static charge correlation 
functions diverge because of the mixing between the charge fluctuations.
Along the spinodal line, the exciton mode remains 
massive (i.e. $\omega_{exc}(q=0) \ne 0$), even if the charge-transfer (CT) 
correlation function is divergent. The instability is instead
accompanied by the overdamped zero sound acquiring a vanishing and 
then negative velocity.
These results are only quantitatively, and not qualitatively, changed if the 
Fock term in Eq. (\ref{fock}) is considered \cite{coul2}. 

When the LR potential of Eq. (\ref{cco}) 
is added to the short-range extended Hubbard model of
Eq. (\ref{short}), it provides a huge electrostatic cost to
the long-wavelength charge fluctuations thereby preventing PS. 
Nevertheless, an instability
region characterized by the divergence of the density-density correlation
functions at a finite $q_c$ is still possible and is indeed present, as in the 
single-band Hubbard-Holstein model,
where an ICDW stripe instability was found \cite{sing,longr}. In Fig. 1 and
(2) we show two phase diagrams for
two different values of the LR Coulomb force
strength ${\tilde V}$: ${\tilde V}=1.6t_{pd}$ and
${\tilde V}=8t_{pd}$ with
$t_{pp}=0.2t_{pd}$ and $V=2.3t_{pd}$ \cite{note3}.
These values of ${\tilde V}$ would 
 correspond to a repulsion between holes on
two nearest-neighbor copper sites of the order $0.1t_{pd}$ and $0.5t_{pd}$)  
respectively. The phase diagrams are determined 
by identifying the divergences of the various correlation functions
within the $1/N$ expansion at the leading order .
The  finite-$q$ instability is indicated by the dashed line.
When the strength of the Coulomb potential ${\tilde V}$
is larger, as in Fig. 2, the unstable region 
shrinks around the  VI  point. 
The dynamical properties of the 
extended Hubbard model at $q=0$
 are not affected by the introduction of the LR forces. A remarkable 
consequence is that
the  VI  point, which can be characterized by
$\chi_{CT}(q=0, \omega\to 0) =\infty$
 occurs for the same $\Delta_{0}$ and $\delta$ of the 
purely short-range case.
In the presence of a (large) 
 $\tilde{V}$, part of the first-order  VI  line 
lays outside the region of the finite-$q$ 
instability. However we have checked that the system 
does not show any critical behavior along this line due to the fact 
the energy of the exciton-mode remains finite on it and no divergences
arise in the response functions as well as in the effective 
scattering amplitude between quasiparticles. 

The static ($\omega=0$) charge-transfer correlation function 
$\chi_{CT}(q,\omega=0)$ is reported in Fig. 3 outside and inside 
the CDW-unstable region, for $\tilde{V}=1.6t_{pd}$. Near
the instability, a huge peak at $q\simeq q_{c}$ develops in the
charge-transfer as well as in all other correlation functions and
they eventually diverge when the charge-instability line is reached. 

Although it may be {\it influenced} by some specific features  of the 
band structure (proximity to a Van Hove singularity, partial nesting
of some portion of the Fermi surface and so on), 
the critical ${\bf q_{c}}$ is not {\it determined} by
any nesting property of the Fermi surface. It rather results from the 
balance between the tendency towards PS 
of the short-range model and the electrostatic cost imposed by 
LR forces to charge segregation. Then its modulus depends both on
doping and, less sensitively, 
 on the Coulomb potential through $\tilde{V}$. It increases 
ranging from $0.5$ to $0.8$ (in units of the inverse of lattice
constant $a$) with doping and, slightly, with $\tilde{V}$. This behavior
reproduces the one already found in the single-band Hubbard-Holstein model,
although in this latter case the value of the modulus of $q_{c}$ was about 
20\% larger. However,  this feature is obviously non-universal and is
a specific outcome of the band parameters considered here.
We found that the correlation functions are rather 
isotropic in the $(q_{x},q_{y})$ space. However a detailed analysis of
the direction along which the instability first occurs
shows that the  $(1,0)$  and $(0,1)$ directions are more favorable 
than the diagonal $(1,1)$ one. 

To clarify the physics underlying the ICDW instability,
we studied the behavior of the collective 
modes at finite $q$ and $\omega$ in the stable region.
To this purpose, we 
analyzed the density-density correlation functions [Eqs. (\ref{kappa}) and
(\ref{ct})].
The frequency and momentum dependence of the poles  (or resonanaces)
of the correlation functions as a function of $q$ give indeed
the dispersion of the collective modes of the model. Two modes can be 
identified: The plasmon mode and the charge-transfer exciton. 
Having introduced
a LR term proportional to $1/q$ (for small $q$) the energy of the 
plasmon mode vanishes at $q=0$
as $\sqrt{q}$, whereas the frequency of the exciton mode at $q=0$ stays finite 
all over the stable region of the phase diagram.
Due to the mixing, both poles are present in all the correlation functions,
although with different spectral weight.  Whereas 
 the peak of the plasmon is more pronounced in
the total density correlation function, the excitonic mode has a
larger weight in the CT correlation function.
  
In Fig. 4 we compare the dispersion of the $2D$-plasmon near 
and far away the CDW instability. 
Fig. 4 clearly shows that the $2D$-plasmon mode softens 
at the same $q$ for which the instability takes place.
On the contrary, the exciton mode remains at higher energies and the CDW
instability does not affect much its dispersion. We have also analyzed
the plasmon mode as a function of the direction of
the momentum $q$ in the Brillouin zone. We find that, up to $q\sim 0.5$ 
(in units of the inverse lattice constant), the dispersion is nearly
isotropic and depends  very weakly on the direction of $q$. 
In particular, when the system is far from 
 the instability, the dispersion on the (1,0) and (1,1) directions is
linear (apart from the small region around $q=0$ where the $q^{1/2}$ 
behavior is found)  with a 10\% higher slope in the (1,1)
direction. This
analysis illustrates that the occurrence of an ICDW instability
in the three-band extended Hubbard model
is accompanied by the softening of the plasmon mode due to
the attractive effective interactions mediated by the charge-transfer
fluctuations in the presence
of a sizable nearest-neighbor repulsion $V$. The microscopic
dynamics of the instability is therefore
substantially different from the analogous instability taking
place in the Hubbard-Holstein model \cite{longr}. In this latter case
the electronic plasma mode mixed with an optic phonon mode.
The finite momentum instability was then accompanied by the 
softening of the phonon mode, which was also responsible for the
effective attraction eventually driving the instability.
However, despite the physically different microscopic
driving forces and the different dynamical evolution of the collective modes,
the similar occurrence of an ICDW instability driven by
the interplay of PS tendency  and LR forces clearly illustrates
the full generality of this mechanism.

\section{Effective interaction}

Our analysis now proceeds with the investigation of the 
effective interaction between quasiparticles on the Fermi 
surface as defined by
\begin{equation}
\label{gamma}
\Gamma(k_{F},k^{\prime}_{F},q;\omega)=
-\sum_{\mu,\nu} \tilde{\Lambda}_{11}^{\mu}(k_{F}^{\prime},-q)
D^{\mu \nu}(q,\omega)
\tilde{\Lambda}_{11}^{\nu}(k_{F},q).
\end{equation}
Fig. 5 displays $\Gamma(q,\omega=0)$ for ${\tilde V}=0.8t_{pd}$ 
and $\delta=0.2$, for two different values of 
$\Delta_{0}$ close to the CDW instability, as a function of 
$(q_x,0)$. As in the case of the short-range model \cite{coul1,coul2,coul3}, 
near the charge instability, a large attractive interaction between
quasiparticles is generated at $\omega=0$. However, while in the 
short-range model the 
interaction diverges negatively at $q=0$, in the presence of LR forces, 
$\Gamma(q,\omega=0)$ stays finite at $q=0$ and diverge negatively
 at $q_{c}$, in correspondence to the divergence of the correlation 
functions. This is
rather natural since at leading order in $1/N$
the singular behavior of both quantities is
related to the singular behavior of the boson propagator.
 
In the proximity of the instability, 
$\Gamma(q,\omega)$ can be fitted by
the following expression
\begin{equation}
\label{effi}
\Gamma(q,\omega)=\tilde{U}
-\frac{\hat{V}}{(q-q_{c})^{2}+\kappa^2-i\gamma \omega}.
\end{equation}
$\tilde{U}$ describes the residual, 
nearly $k$-independent repulsion mediated by the slave bosons: 
 within
the large-N slave-boson formalism, the infinite U repulsion between the 
bare fermions, is reduced to a rather weak
residual repulsion between the Fermi-liquid quasiparticles.
The imaginary term in the denominator of Eq. (\ref{effi}) is proportional
to $\omega$ through the  damping coefficient $\gamma$ and
reproduces the behavior
of the imaginary part of the bare density-density
polarization bubble of the quasiparticles $Im\chi_{nn}^0(\omega,q)$
at small frequencies and finite momenta
close to $q_c$. This means that, despite the complicated structure 
of the scattering amplitude Eq. (\ref{gamma}), in the proximity of 
the charge instability, the interaction between quasiparticles has a 
simple RPA-like form
\begin{equation}
\label{effirpa}
\Gamma(q,\omega)=\tilde{U}
-\frac{\hat{V}}{1+\Gamma_{\omega}(q, \omega)\chi_{nn}^0(q,\omega)}
\end{equation}
with $\Gamma_\omega (q, \omega)$ being an effective dynamical
interaction between the quasiparticles. $\Gamma_{\omega}(q, \omega)$
is dominantly attractive at $q\sim q_c$ and is
 mediated by virtual high-energy
processes (mostly interband transitions). 
The behavior of $\kappa^2$
as a function of $\delta-\delta_{c}$ is plotted
in Fig. 6. It vanishes as $\alpha (\delta-\delta_{c})^{2\nu}$ with 
$2\nu=1$. On general grounds, this is what one expects approaching a 
Gaussian QCP. Since the leading-order $1/N$ expansion bears
resemblance with a RPA resummation, it is quite natural that 
the propagator of the critical fluctuations assumes the universal form
of Eq. (\ref{effi}), which is also found in the
context of the Hubbard-Holstein model \cite{sing,longr} and in the
proximity of an antiferromagnetic QCP  
 \cite{MMP}, where the instability also occurs for
a finite value of $Q_{AF}=(\pi,\pi)$. Of course, the specific form
of Eq. (\ref{effi}) could depend from our approximate
(nearly mean-field) treatment. Nevertheless the singular
nature of the interactions mediated by critical fluctuations
is a sound generic consequence. Many physical consequences
stem from the presence of a QCP related to an ICDW instability,
which have already been generically explored in previous works
\cite{sing,perali}. In particular, the singular interactions were 
related to an anomalously large decay ratio for the quasiparticles
at the ``hot'' spots [$1/\tau \sim \sqrt{max[\epsilon_k,T]}$],
to linear-in-T resistivity in $2D$ and $\rho \sim T^{3/2}$ in $3D$
\cite{sing} and a specific doping dependence of the superconducting
critical temperature $T_c$ was found \cite{perali}.
Similar results are obviously obtained, since 
the above analyses are quite general in so far they are 
based on the general theory of QCP's. On the other hand,
the peculiar character of the presently considered model obviously
enter in the values of the non-universal constants.
In particular, whereas the temperature dependence of the mass
parameter $\kappa^2$ in Eq. (\ref{effi}),
can only be calculated within a finite-temperature analysis,
other parameters are accessible within our $T=0$ slave-boson
calculation. Specifically, as seen in Fig.6, 
for typical Hamiltonian parameter, the zero-temperature
doping dependence of the mass $\kappa^2=\alpha(\delta-\delta_c)$ 
is determined by the value of 
the coefficient $\alpha \sim 2$ and turns out to be
quite smaller than in the phononic Hubbard-Holstein
model.  As a consequence in the present case the mass term 
 grows more slowly upon increasing the doping away from 
$\delta_c$, thus leading to a more extended region with
substantial critical fluctuations.

\section{Discussion and Conclusions}

The scenario here presented in the context of the 
three-band Hubbard model involves the relevant 
interplay between the plasma and the charge-transfer
modes. An obvious completion of our work would
then regard the possible experimental confirmation
of the physical outcomes of the model. Unfortunately,
such a connection with experiments is not presently
attainable on a quantitative level, due to
the lack of detailed experimental analyses on
collective electronic excitations at finite momenta.

As far as the limitations of our model are concerned,
these are manyfold. First of all, we only considered
a purely two-dimensional electron system embedded
in a $3D$ inert screening medium, thus neglecting the 
layered nature of the superconducting cuprates.
As a consequence, the investigated collective modes only have
a purely two-dimensional character.  It is well
known that, as far as the $2D$ plasma mode is concerned,
it qualitatively differs from the collective plasma
modes of a layered electron gas (LEG) \cite{leg}
formed by an array of $2D$ electron gas layers
spaced by a distance $d$. In particular, the screening 
induced by the other layers changes the
$\sqrt{q}$ dispersion of the $2D$ mode into a 
continuous branch of modes labelled by the wavevector
$q_z$ in the direction perpendicular to the layers,
with $-(\pi /d) \le q_z \le \pi /d$. Whereas the $k_z=0$
mode is massive like in $3D$ isotropic systems
(plasma oscillations in phase on all different layers), 
the other modes
are acustic with different velocities ($v \to \infty$ for
$k_z \to 0$) until at finite momenta 
in the $2D$ Brillouin zone they all merge in a
narrow, infinitely degenerate branch practically 
indistinguishable from the dispersion of the purely
$2D$ mode. However, it is our opinion that 
some important qualitative indications can already
be gained from our simplified model, since
the LEG modes merge at $\tilde{q} \sim \pi/d$
substantially smaller than the typical momenta $q$ of the
twodimensional Brillouin zone. Therefore,
although explicit calculations are still
in progress, we believe that the physics at finite
sizable momenta of the twodimensional Brillouin zone
is not very different for the isolated or the 3D layered
systems \cite{HEQ} and a CDW instability is a natural outcome
provided $q_{c}$ is larger than $\tilde{q}$. In our case,
we found,  $q_{c} \sim 0.5$
(remember that we take the lattice spacing $a=1$)
thus restricting the validity of our analysis to systems
where $\tilde{q} \lesssim 0.5$, i.e. $d >6 a$.

Other important simplifications of the model here considered
are worth being investigated and could prevent a direct
 comparison with real systems. Specifically in the 
present model, we did not include a small
 hybridization ($t_{\perp}$) between adjacent layers,
which would render the plasma modes 
with $k_z \ne 0$ (weakly) massive, thus modifing the
dynamics at small momenta. Furthermore the interplay
between the mixed collective modes and the particle-hole
excitations was  considered within a $1/N$ expansion
bearing a close resemblance with the RPA approximation.
Although extensively used in the literature, this approximation
is not guarantied to provide a quantitative description
of the physics of the strongly interacting electrons.
In particular, at the leading order in $1/N$, both the screening
and the damping of the collective modes is determined by the
quasiparticles, without any account for the incoherent
excitations leading to incoherent particle-hole continua
on much larger energy scales. These excitations could in principle
affect the dynamics of the system by providing a damping
channel for the collective modes even when these latters
are outside the quasiparticle-quasihole continuum.

Besides the above  limitations of the
theoretical model, a
comparison with experiments is also made difficult
by  the lack of detailed experimental analyses
of electronic excitations at finite momenta. 
In particular the extensive wealth of data
provided by neutron scattering experiments
could be related to the above discussed physical effects
only in the case of a strong enough coupling of the
CT and plasma collective modes with the lattice or 
spin degrees of freedom. This could render too indirect the
access to the physics of the electronic collective modes,
with probably too little intensity to get detectable effects.

On the other hand a direct
access to electronic excitations at finite momenta
is given by electron-energy-loss spectroscopy.
However, the presently available analyses, do not seem to us conclusive
in order to confirm or to disprove any anomalous dynamical
behavior of the collective plasma modes. In particular, our analysis
showed (see Section III) that the anomalous softening of the
plasma modes at $q_c$ occurs in a rather narrow range of doping
$\delta \sim \delta_c \sim \delta_{opt}$ and momenta $q\sim q_c$.
This by itself would require a rather detailed and dedicated search.
Moreover, it is to be remembered that (local) superconducting pairing
``quenches'' the ICDW instability, thereby preventing a complete 
softening of the plasma mode. Then, presumably the only consequence
of the unachieved charge instability would be a partial softening
of the mode, likely accompanied by an increase of incoherent 
spectral weight at low energies near the critical $q_c$. This
latter effect should be separated from the usual substantial
background, which likely will render its detection  quite
a difficult task. 

Despite the above theoretical and experimental limitations,
the present work provides a definite contribution to the
theoretical substantiation of the ICDW-QCP scenario. 
Indeed we proved that LR coulombic forces do not completely
stabilize the tendency towards charge instability
in the three-band Hubbard model extended with nearest-neighbour Coulomb 
repulsion. In particular we explicitely demonstrated the 
existence of an ICDW instability for a substantial parameter
range of the model and for realistic doping values. Moreover,
we showed that, although the VI is still present
in the phase diagram and remarkably the excitonic softening
at $q=0$ at the VI critical point is not influenced by
the LR forces, the ICDW instability is dominant 
as it embeds the VI point thus providing the only possible
source of critical singular scattering. 

Although we have presented the results in the strong-coupling limit,
the qualitative behavior of the model does not change in the weak-coupling
case (we report the calculations in Appendix B). We have also checked that 
both in the strong- and the 
weak-coupling approach the inclusion of a Fock term does
not change qualitatively the properties of the system: The CDW region is
still present, surrounding the  VI  point and no critical scattering is found
on the  VI  line. 

These results support the idea that 
 incommensurate CDW is a common feature of
correlated-electron systems irrespective of the
microscopic interaction mechanisms. 

Indeed a similar 
scenario was also found in the framework of one-band 
Hubbard model with phononic interactions
\cite{longr}. 
As far as the collective modes are concerned, their dynamical behavior
is obviously related to the specific interactions:
In the present case the instability is characterized by the softening of 
the plasma mode, whereas in the Hubbard-Holstein model the CDW 
instability is accompanied by the softening of the phonon mode, leading to an 
instability of the underlying lattice structure. 
It is quite remarkable that such different dynamics of the collective
excitations of the systems, reflecting the underlying microscopic 
differences of the two models, eventually lead to the same
generic conclusion as far as the existence of the ICDW critical point
is concerned and the concomitant presence of singular scattering.
Indeed, the effective interaction
is found in both models to have the same singular behavior as a function of
$\delta-\delta_c$ and to display an imaginary part linearly dependent on
the frequency. In both models, 
the singular effective interaction is found to be attractive in the
particle-hole channel for a sizeable range of $q$ around $q_{c}$
in the stable region close to the CDW instability.  
The scenario presented so far does not include superconductivity.
However, it is rather obvious (and it was proven directly in the
Hubbard-Holstein model \cite{longr}) that the 
strong effective attraction between quasiparticles,
also shows up in the particle-particle channel 
and leads to (local) Cooper pairing. The occurrence
of superconductivity (or of superconducting pairs without long-range
coherence) would definitely modify the above scenario in so far
it provides an alternative to ICDW. 
A true long-range CDW order is actually
realized overcoming the quenching pairing
tendency only when a commensurability condition is
realized which pins the charge fluctuations.
When commensurability effects do not occur,
ICDW fluctuations and superconducting pairing 
interplay and compete, with strong effects also on the
magnetic response of the system. On the one hand ICDW
fluctuations create dynamical charge depleted stripes, where
magnetic correlations may subsist even near optimal doping. 
On the other hand the (local) Cooper pairing induced by the singular ICDW
scattering can be responsible for the charge- and spin-gap effects
which arise in the underdoped cuprates.
This complex interplay of ICDW, Cooper pairing and magnetism
is not of our
concern here, but is surely a most interesting (and difficult) subject
which is relevant for the 
understanding of the cuprates \cite{cina,evora}.

\acknowledgments
This work was supported by the Istituto Nazionale di Fisica della
Materia with the Progetto di Ricerca Avanzata 1996.

\appendix

\section{Boson-fermion vertices}

In this section we report the boson-fermion vertices in the orbital basis 
which are needed in the hamiltonian defined in Eq. (\ref{int}):
\begin{equation}
\Lambda^{1}(k,q)=-2r_{0}t_{tp} 
\left (
\begin{array}{ccc}
0                                                    & 
\sin \left (\frac{k_{x}-\frac{q_{x}}{2}}{2} \right ) &
\sin \left (\frac{k_{y}-\frac{q_{y}}{2}}{2} \right ) \\
\sin \left (\frac{k_{x}+\frac{q_{x}}{2}}{2} \right ) & 
0                                                    &
0                                                    \\
\sin \left (\frac{k_{y}+\frac{q_{y}}{2}}{2} \right ) & 
0                                                    &
0
\end{array}
\right ),
\end{equation}
\begin{equation}
\Lambda^{2}(k,q)= 
\left (
\begin{array}{ccc}
i & 0 & 0 \\
0 & 0 & 0 \\
0 & 0 & 0
\end{array}
\right ),
\;\;\;
\Lambda^{3}(k,q)= 
\left (
\begin{array}{ccc}
1 & 0                                      & 
0                                          \\
0 & -\cos \left ( \frac{q_{x}}{2} \right ) & 
0                                          \\
0 & 0                                      &
-\cos \left ( \frac{q_{y}}{2} \right ) 
\end{array}
\right ),
\end{equation}
\begin{equation}
\Lambda^{4}(k,q)= 
\left (
\begin{array}{ccc}
i & 0                                      & 
0                                          \\
0 & i\cos \left ( \frac{q_{x}}{2} \right ) & 
0                                          \\
0 & 0                                      &
i\cos \left ( \frac{q_{y}}{2} \right ) 
\end{array}
\right ),
\;\;\;
\Lambda^{5}(k,q)= 
\left (
\begin{array}{ccc}
i & 0 & 0 \\
0 & 0 & 0 \\
0 & 0 & 0
\end{array}
\right ),  \;\;\;
\Lambda^{6}(k,q)= 
\left (
\begin{array}{ccc}
0 & 0 & 0 \\
0 & i & 0 \\
0 & 0 & 0
\end{array}
\right ), 
\;\;\;
\Lambda^{7}(k,q)= 
\left (
\begin{array}{ccc}
0 & 0 & 0 \\
0 & 0 & 0 \\
0 & 0 & i
\end{array}
\right ).
\end{equation}

\section{Two-band model in weak-coupling approach}

In this Appendix, we present an analytic treatment 
of a simplified model, showing 
that (i) the energy of the exciton mode at  $q=0$
(denoted by $\omega_{exc}$) is not affected by LR interactions so that 
the softening of this mode (VI  point) takes place in the same point of 
the phase diagram and that (ii), once the pure SR system is unstable 
towards phase separation, 
the system with LR forces always displays a finite-$q$ 
instability for any value of ${\tilde V}$. 
In order to keep the formal structure more transparent
we confine the calculations to the weak-coupling case \cite{notasc}.

The main difference with respect to the strong-coupling case is
given by the absence of the slave-bosons $r_{q}$ and $\lambda_{q}$ needed to
treat the infinite on-site repulsion. 
To further simplify the treatment we also neglect 
the direct oxigen-oxigen overlap ($t_{pp}=0$).
In this case one combination of oxygen orbitals 
does not hybridize with the copper $d$ orbitals, and gives rise to 
a flat non-bonding band. Interband processes involving this
non-bonding band are decoupled in the small-$q$ limit and, even
at finite momenta, do not play any qualitatively relevant role.
Therefore, for the sake of simplicity, we will only consider
the simplified  two-band Hubbard model. 

The form factor 
$\gamma_{k}=\sqrt{\sin^{2} \left ({k_{x}}/{2} \right )+
\sin^{2} \left ({k_{y}}/{2} \right ) }$
can be introduced to relate the Fourier transform $p_{k}$ of 
the bonding oxygen orbital combination $p_{i}$, to the Fourier trasform
of the copper $d$ orbitals. The resulting
mean-field Hamiltonian is given by:
\begin{equation}
\label{twoband}
H_{MF}=\sum_{k,\sigma}{\cal H}_{MF}(k)=
\left (
\begin{array}{cc}
p_{k}^{\dag} & d_{k}^{\dag} 
\end{array}
\right ) 
\left (
\begin{array}{cc}
\epsilon_{p }                 & -\sqrt{2}t_{pd} \gamma_{k} \\
   -\sqrt{2}t_{pd} \gamma_{k} & \epsilon_{d}
\end{array}
\right )
\left (
\begin{array}{c}
p_{k}  \\
d_{k} 
\end{array}
\right ),
\end{equation}
where $\epsilon_{d}$ and $\epsilon_{p}$ are
the atomic levels, including the Hartree shifts. 
The mean-field hamiltonian in Eq. (\ref{twoband}) can be diagonalized by a 
standard unitary transformation, giving the two bands
$E_\pm (k)=\left[\epsilon_{p}+\epsilon_{d}\pm
\sqrt{\Delta^2+8t_{pd}^2\gamma_k^2} \right]/2$.
We define $\Delta \equiv \epsilon_{p}-\epsilon_{d}$.

We are interested in the analysis of the physical response functions
${\hat{\chi}}_{\pm,\pm}(q,\omega)$
for the total density $(+)$ and CT $(-)$ fluctuations.
A simple analytic progress can only be made in the small-$q$ limit, to which
we will confine our treatment here. In this limit the complicated
internal structure of the LR interaction 
[see Eqs. (\ref{cco})-(\ref{cupo})] greatly simplifies,
and the cosine form factors in the nearest-neighbor
interaction [see the expressions of the $\Lambda^{3,4}$ vertices
in Appendix A] drop. Moreover, in the spirit of weak-coupling 
approach, we treat the interaction terms in Eq. (\ref{lamadreditutto}) 
by means of 
a standard Hartree decoupling. 
(We will also show below that adding the contribution of the 
Fock decoupling does not change qualitatively the behavior of the model).

The Hartree decoupling of the n-n coulomb repulsion is the following
\begin{equation}
V \sum_{i \sigma \sigma^{\prime}} 
d_{i \sigma}^{\dag}d_{i \sigma} 
p_{i \sigma^{\prime}}^{\dag}p_{i \sigma^{\prime}}=
V \sum_{i \sigma} 
\left [ 
2n_{i}^{d} p_{i \sigma}^{\dag}p_{i \sigma}+
2n_{i}^{p} d_{i \sigma}^{\dag}d_{i \sigma} \right ] - 8V n^{p}n^{d} 
\label{hartee}
\end{equation}
where $n_{i}^{d}$ and $n_{i}^{p}$ are the local values of the copper 
and oxygen density per spin. 

The resulting interaction matrix (in the
total-density and CT basis) reads
\begin{equation}
\hat{V}(q)=
\left (
\begin{array}{cc}
-2V - U/4 - V_c(q)/2 & -U/2 \\
-U/2 & 2V - U/4 
\end{array}
\right ).
\end{equation}
with $V_c(q)\approx \tilde{V}/q$
and where the local U interaction on copper
was also introduced. At gaussian level, the 
response functions are given by the matrix relation
\begin{equation}
\label{gaussian}
\hat{\chi}(q,\omega)=-\frac{\hat{\Pi}_{0}(q,\omega)}
{\hat{1}+\hat{\Pi}_{0}(q,\omega) \hat{V}(q)}.
\end{equation}
involving the 
bare-bubble matrix:
\begin{equation}
\label{2bl}
\hat{\Pi}_{0}^{\alpha \beta}(q,\omega)=\sum_{k,\nu} Tr \left \{ 
\hat{\Lambda}^{\alpha}(k+q) \hat{G}^{orb}(k+q,\omega+\nu)
\hat{\Lambda}^{\beta}(k) \hat{G}^{orb}(k,\nu) \right \},
\end{equation}
where $\hat{G}^{orb}_{\alpha,\beta}(k,\nu)$ 
is the matrix of the Fermionic Green functions 
in the orbital basis $(\alpha,\beta = d,p)$
and $\hat{\Lambda}^{\alpha}(k)$ are the vertices, which couple the 
$p$ and $d$ fermions to the density fluctuations
\begin{equation}
\hat{\Lambda}^{+}(k)=
\left (
\begin{array}{cc}
1 & 0 \\
0 & 1
\end{array}
\right ),\;\;\;
\hat{\Lambda}^{-}(k)=
\left (
\begin{array}{cc}
1 & 0  \\
0 & -1
\end{array}
\right ),
\end{equation}
The value of the CT exciton frequency at zero momentum can 
be extracted from the pole of the CT correlation function, since,
due to particle conservation, the total density ($+$) decouples from the
dynamics. This shows up in the $q \rightarrow 0$, $\omega \neq 0$ 
behavior of the bubbles containing the total-density vertex:
\begin{equation}
\lim_{q \rightarrow 0}\Pi_{0}^{+\alpha}(q,\omega) \sim q^{2},
\end{equation}
($\alpha=+,-$). 
Then, since $V_{c}(q)\sim \tilde{V}/q$ always couples to
a total-density vertex of $\Pi_{0}^{+\alpha}$, the diverging
Coulomb potential is cancelled by the 
small-$q$ behavior of the bare bubbles having at least
one $(+)$ vertex and does not appear in
the $q=0$ limit of Eq. (\ref{gaussian}).
By introducing the notation
$$
\Pi_{0\omega}^{\mu \nu}=\lim_{q \rightarrow 0} \Pi_{0}^{\mu \nu}(q,\omega)
~~~~~~~~~ 
\Pi_{0q}^{\mu \nu}=\lim_{\omega \rightarrow 0} 
\Pi_{0}^{\mu \nu}(q \simeq 0,\omega),
$$
one obtains
\begin{equation}
\chi_{CT}(0,\omega)=\chi_{--}(0,\omega)=\frac{\Pi_{0\omega}^{--}}
{\omega^{2}- \left ( 1+\Pi_{0\omega}^{--}V_{--} \right )}.
\label{dinweak}
\end{equation}
Notice that $V_c$ no longer appears in Eq. (\ref{dinweak}), thus showing
that $\omega_{exc}(q=0)=
\left ( 1+\Pi_{0\omega}^{--}V_{--} \right)^{1/2}$ is the same as 
in the pure SR case.

In the opposite static limit $\omega=0$ and small $q$
\begin{equation}
\chi_{--}(q)=\chi_{CT}(q)=
-\frac{ \left \{ \Pi_{0q}^{++} \Pi_{0q}^{--}-
\left (\Pi_{0q}^{+-} \right )^{2} \right \} V^{++}}
{\left \{ \Pi_{0q}^{++}+ 
V^{--} \left [ \Pi_{0q}^{++} \Pi_{0q}^{--}- \left (\Pi_{0q}^{+-} \right )^{2}
\right ] \right \} V^{++}+A^{2}_{H}}.
\label{statweak}
\end{equation}
with
\begin{equation}
A^{2}_{H}= 1+V^{--}\Pi_{0q}^{--}+2V^{+-}\Pi_{0q}^{+-}+\left(V^{+-}\right)^2
\left[ \left(\Pi_{0q}^{+-}\right)^2 - \Pi_{0q}^{++}\Pi_{0q}^{--}\right] > 0.
\end{equation}
Using the relation 
\begin{equation}
\lim_{\omega \rightarrow 0} \Pi_{0\omega}^{--}=
\lim_{q \rightarrow 0} \left[ \Pi_{0q}^{--}-
\frac{\left (\Pi_{0q}^{+-} \right )^{2}}{\Pi_{0q}^{++}}\right]
\end{equation}
and noting that, for small $q$'s, $V^{++}$ is dominated by the
LR coulombic term $\tilde{V}/q$,
a simpler form can be obtained for $\chi_{--}(q)$
\begin{equation}
\label{smallq}
\chi_{--}(q)\approx {\tilde{V}\Pi_{0\omega}^{--}
\over 
\left(\omega_{exc}^2+B^2q^2 \right)\tilde{V}+q
\left(A_H^2/\Pi_{0q}^{++}\right)}
\end{equation}
It worth noting that, due to this relations,
the static limit of the correlation function equals the dynamical limit,
provided $\omega_{exc}$ is finite. This fact depends on the presence of 
the LR forces: in the dynamical limit
the intra-band processes are switched off and only inter-band 
transition are allowed; 
in the static limit both processes are allowed, but the LR interactions
drastically reduces the intra-band transitions at $q=0$, while it allows
the inter-band ones. This also justifies why the exciton energy does 
not change introducing LR forces: $\omega_{exc}$ is indeed involved in 
inter-band processes which are not affected.

The zeros of the denominator of Eq. (\ref{smallq}) can be 
obtained through the equation
\begin{equation}
\label{equati}
{\tilde V} \left ( \omega_{exc}^{2}+B^{2}q^{2} \right )-A^{2}q=0.
\end{equation}
with $A^2\equiv A_H^2/\Pi_{0q}^{++}$.
If $\omega_{exc}<A^{2}/B{\tilde V}$, 
Eq. (\ref{equati}) has two real solutions showing the instaurance of a 
finite-$q$ instability
\begin{equation}
\label{qc}
q_1 \simeq \frac{\omega_{exc}^{2} {\tilde V}^{2}}{A^{2}}~~~~~~q_2 
\simeq \frac{A^{2}}{{\tilde V}B^{2}},
\end{equation}
Approaching the  VI  point upon changing the doping, $\omega_{exc}$
decrease. Eventually it reaches the {\it finite}
critical value $\omega_{exc}^c$ where
the charge-transfer (and any other) correlation function diverges at 
finite-$q$ and the instability takes place.\\
From the above result we see that there is a direct connection between the
lowering of the exciton energy $\omega_{exc}$ and the developing of a
region of instability with the finite-$q$ softening of the plasmon mode.
A similar behavior can be found in the SR model \cite{coul1}, where the $q=0$ 
instability is driven by the partial softening of 
the exciton mode which pushes the 
zero-sound into the continuum eventually leading the system to a
phase-separation instability.

Finally, we show that the inclusion of a Fock contribution does not change
the qualitative behavior of the model. For the sake of simplicity we limit 
ourself to the weak-coupling approach, but the same conclusions hold also in 
the strong-coupling case.
In this case we have to add to Eq. (\ref{hartee}) the following terms:
\begin{equation}
V \sum_{i\sigma} \left(
Z_{i} d_{i \sigma}^{\dag}p_{i \sigma}+
Z_{i}^{\dag} p_{i \sigma}^{\dag}d_{i \sigma} \right)
-2V Z^{\dag}Z 
\end{equation}
We introduce two suitable linear combinations of the Fock bosons:
\begin{eqnarray}
A_{i}=\frac{1}{2}\left (Z_{i}+Z_{i}^{\dag} \right ) \nonumber\\
B_{i}=\frac{i}{2}\left (Z_{i}-Z_{i}^{\dag} \right ) \nonumber
\end{eqnarray}
Considering a mean-field and fluctuating part for these bosons 
($A_{i}=A_{0}(1+\delta A_{i})$, $B_{i}=B_{0}(1+\delta B_{i})$), 
two new vertex are needed in the bare-density-correlation functions (we note
in passing that the $A$-vertex has the same structure of the $r$-vertex in the 
strong-coupling approach) :
\begin{equation}
\hat{\Lambda}^{A}(k)=
\left (
\begin{array}{cc}
0                            & - 2 V A_0 \gamma_{k} \\
-2 V A_0 \gamma_{k} & 0
\end{array}
\right ), \;\;\;
\hat{\Lambda}^{B}(k)=
\left (
\begin{array}{cc}
0                            & 2 V  B_0 \gamma_{k} \\
-2 V B_0 \gamma_{k} & 0
\end{array}
\right ).
\end{equation}
Due to the fact that the boson $B_{i}$ decouples from the others we can 
easily write the dynamical limit of the correlation function 
$\chi_{CT}(q)$ including the Fock terms:
\begin{equation}
\chi_{CT}(\omega)=\chi_{--}(\omega)=\frac{\Pi_{0\omega}^{--}}
{\omega^2-\left( 1+\Pi_{0\omega}^{--}V_{--}+\Pi_{0\omega}^{AA}V_{AA}\right) },
\end{equation}
where $V_{AA}=-2V$. This result
is the same as the one obtained without the Fock term apart 
from a contribution
proportional to $V_{AA}$ which simply renormalizes the exciton-mode
energy $\omega_{exc}^{HF}$ \cite{coul2}.

Furthermore the statical limit is:
\begin{equation}
\chi_{--}(q)=\chi_{CT}(q)=
-\frac{ \left \{ \Pi_{0q}^{++} \Pi_{0q}^{--}-
\left (\Pi_{0q}^{+-} \right )^{2} \right \} V^{++}}
{\left \{ \Pi_{0q}^{++}+ 
V^{--} \left [ \Pi_{0q}^{++} \Pi_{0q}^{--}- \left (\Pi_{0q}^{+-} \right )^{2}
\right ] 
+ V_{AA} \left [ \Pi_{0q}^{++} \Pi_{0q}^{AA}- \left (\Pi_{0q}^{+A} \right )^{2}
\right ] \right \} V^{++}+A^{2}_{HF}}.
\label{statweakfock}
\end{equation}
This expression can be cast in the same form of Eq. (\ref{statweak}) provided
$\omega_{exc}$ is replaced by $\omega_{exc}^{HF}$ and so
all the conclusions we have drawn after Eq. (\ref{smallq})
still hold.

{\bf Figure Captions}

{Fig. 1: Phase diagram in the plane 
$\epsilon_{p}^{o}-\epsilon_{d}^{o}$ vs $\delta$ for 
$t_{pp}=0.2 t_{pd}$, $V=2.3 t_{pd}$. The Coulomb interaction strenght is
taken so that the repulsion between two nearest-neighbor copper atoms, 
$V_c(d-d)$, is $0.1 t_{pd}$.}

{Fig. 2: Phase diagram in the plane $\epsilon_{p}^{o}-\epsilon_{d}^{o}$ 
vs $\delta$ for 
$t_{pp}=0.2 t_{pd}$, $V=2.3 t_{pd}$. The Coulomb interaction strenght is
taken so that the repulsion between two nearest-neighbor copper atoms, $V_c(d-
d)$, is $0.5 t_{pd}$.}

{Fig. 3: Charge-Transfer correlation function $\chi_{CT}(q,0)$ vs momentum
for $\delta=0.12$, $\tilde{V}=1.6t_{pd}$
 and $\epsilon_{p}^{o}-\epsilon_{d}^{o}=1.835t_{pd}$ (+), 
$\epsilon_{p}^{o}-\epsilon_{d}^{o}=1.84t_{pd}$ ($\Box$), 
$\epsilon_{p}^{o}-\epsilon_{d}^{o}=1.845t_{pd}$ ($\times$), 
$\epsilon_{p}^{o}-\epsilon_{d}^{o}=1.85t_{pd}$ ($\diamond$). 
The critical value is $\epsilon_{p}^{o}-\epsilon_{d}^{o}\sim 1.85t_{pd}$.}

{Fig. 4: Plasma frequency $\omega_{pl}$ vs momentum for $\delta=0.12$ and
$\epsilon_{p}^{o}-\epsilon_{d}^{o}=1.83t_{pd}$ ($\diamond$), 
$\epsilon_{p}^{o}-\epsilon_{d}^{o}=1.845t_{pd}$ (+); the
critical value is $\epsilon_{p}^{o}-\epsilon_{d}^{o}\sim 1.85t_{pd}$}

{Fig. 5: Effective Interaction $\Gamma(q,0)$ vs momentum
for $\delta=0.2$ and $\epsilon_{p}^{o}-\epsilon_{d}^{o}=
1.92 t_{pd}$ ($\diamond$), 
$\epsilon_{p}^{o}-\epsilon_{d}^{o}
=1.922t_{pd}$ (+). The CDW instability corresponds to
$\epsilon_{p}^{o}-\epsilon_{d}^{o} \sim 1.923 t_{pd}$}

{Fig. 6: Doping dependence of the mass parameter at $T=0$
near the critical point. 



\begin{references}
\bibitem   {note3} We have checked that
different values of the nearest-neighbour Coulomb term (greater than $V^{*}$)
do not give different results from the qualitative point of view.
\bibitem {notasc} The strong-coupling extension of the 
weak-coupling treatment is rather straightforward, since,
besides the total density and  the
CT fluctuations,  only one
additional bosonic field, $r_i$, is coupled to the
fermionic fluctuations, thus giving rise to additional response
fuctions. All other fields, $\lambda_i$, $X_i$,
$Y_i$, and $\Omega_i$, couple to the total density or to the
CT density fluctuations (or to their combinations), already 
considered in the weak-coupling approach presented below.
\end{references}
\end{document}